\PassOptionsToPackage{unicode}{hyperref}
\PassOptionsToPackage{hyphens}{url}
\documentclass[
]{article}
\usepackage{xcolor}
\usepackage{amsmath,amssymb}
\usepackage{iftex}
\ifPDFTeX
  \usepackage[T1]{fontenc}
  \usepackage[utf8]{inputenc}
  \usepackage{textcomp} % provide euro and other symbols
\else % if luatex or xetex
  \usepackage{unicode-math} % this also loads fontspec
  \defaultfontfeatures{Scale=MatchLowercase}
  \defaultfontfeatures[\rmfamily]{Ligatures=TeX,Scale=1}
\fi
\usepackage{lmodern}
\ifPDFTeX\else
\fi
\IfFileExists{upquote.sty}{\usepackage{upquote}}{}
\IfFileExists{microtype.sty}{% use microtype if available
  \usepackage[]{microtype}
  \UseMicrotypeSet[protrusion]{basicmath} % disable protrusion for tt fonts
}{}
\makeatletter
\@ifundefined{KOMAClassName}{% if non-KOMA class
  \IfFileExists{parskip.sty}{%
    \usepackage{parskip}
  }{% else
    \setlength{\parindent}{0pt}
    \setlength{\parskip}{6pt plus 2pt minus 1pt}}
}{% if KOMA class
  \KOMAoptions{parskip=half}}
\makeatother
\usepackage{longtable,booktabs,array}
\usepackage{calc} % for calculating minipage widths
\usepackage{etoolbox}
\makeatletter
\patchcmd\longtable{\par}{\if@noskipsec\mbox{}\fi\par}{}{}
\makeatother
\IfFileExists{footnotehyper.sty}{\usepackage{footnotehyper}}{\usepackage{footnote}}
\makesavenoteenv{longtable}
\usepackage{graphicx}
\makeatletter
\newsavebox\pandoc@box
\newcommand*\pandocbounded[1]{% scales image to fit in text height/width
  \sbox\pandoc@box{#1}%
  \Gscale@div\@tempa{\textheight}{\dimexpr\ht\pandoc@box+\dp\pandoc@box\relax}%
  \Gscale@div\@tempb{\linewidth}{\wd\pandoc@box}%
  \ifdim\@tempb\p@<\@tempa\p@\let\@tempa\@tempb\fi% select the smaller of both
  \ifdim\@tempa\p@<\p@\scalebox{\@tempa}{\usebox\pandoc@box}%
  \else\usebox{\pandoc@box}%
  \fi%
}
\def\fps@figure{htbp}
\makeatother
\ifLuaTeX
  \usepackage{luacolor}
  \usepackage[soul]{lua-ul}
\else
  \usepackage{soul}
\fi
\usepackage{bookmark}
\IfFileExists{xurl.sty}{\usepackage{xurl}}{} % add URL line breaks if available
\title{Measuring Research
Convergence in Interdisciplinary Teams Using Large Language Models and
Graph Analytics}
\author{
\small
Wenwen Li\textsuperscript{1},
Yuanyuan Tian\textsuperscript{1},
Sizhe Wang\textsuperscript{1},
Amber Wutich\textsuperscript{2}, 
Paul Westerhoff\textsuperscript{3},\\
\small
Sarah Porter\textsuperscript{4},
Anais Roque\textsuperscript{5},
Jobayer Hossain\textsuperscript{2},
Patrick Thomson\textsuperscript{2},
Rhett Larson\textsuperscript{6},\\
\small
Michael Hanemann\textsuperscript{7}
}

\date{}

\begin{document}
\maketitle

\small
\textsuperscript{1} School of Geographical Sciences and Urban Planning,
Arizona State University, Tempe, AZ

\textsuperscript{2} School of Human Evolution and Social Change, Arizona
State University, Tempe, AZ

\textsuperscript{3} School of Sustainable Engineering and the Built
Environment, Arizona State University, Tempe, AZ

\textsuperscript{4} Kyl Center for Water Policy, Arizona State
University, Tempe, AZ

\textsuperscript{5} Nicholas School of the Environment, Duke University,
Durham, NC

\textsuperscript{6} Sandra Day O\textquotesingle Connor College of Law,
Arizona State University, Tempe, AZ

\textsuperscript{7} Center for Environmental Economics and
Sustainability Policy, School of Sustainability, Arizona State
University, Tempe, AZ

\vspace{12pt}

Correspondence: \href{mailto:wenwen@asu.edu}{\ul{wenwen@asu.edu}}

\vspace{12pt}

\textbf{Abstract}: Understanding how interdisciplinary research teams
converge on shared knowledge is a persistent challenge. This paper
presents a novel, multi-layer, AI-driven analytical framework for
mapping research convergence in interdisciplinary teams. The framework
integrates large language models (LLMs), graph-based visualization and
analytics, and human-in-the-loop evaluation to examine how research
viewpoints are shared, influenced, and integrated over time. LLMs are
used to extract structured viewpoints aligned with the
\emph{Needs-Approach-Benefits-Competition (NABC)} framework and to infer
potential viewpoint flows across presenters, forming a common semantic
foundation for three complementary analyses: (1) similarity-based
qualitative analysis to identify two key types of viewpoints, popular
and unique, for building convergence, (2) quantitative cross-domain
influence analysis using network centrality measures, and (3) temporal
viewpoint flow analysis to capture convergence dynamics. To address
uncertainty in LLM-based inference, the framework incorporates expert
validation through structured surveys and cross-layer consistency
checks. A case study on water insecurity in underserved communities as
part of the Arizona Water Innovation Initiatives demonstrates increasing
viewpoint convergence and domain-specific influence patterns,
illustrating the value of the proposed AI-enabled approach for research
convergence analysis.


\section{\texorpdfstring{\textbf{1. Introduction}
}{1. Introduction }}\label{introduction}

In recent years, the interdisciplinary nature of scientific research has
advanced into a convergence research paradigm (Petersen et al. 2023).
Convergence research emerges as an expanded form of interdisciplinary
research (Finn et al. 2022), integrating knowledge, insights,
methodologies, tools, and mindsets from diverse disciplines through
collaboration. This research paradigm can produce innovative solutions
for complex real-world problems that cannot be solved when considered
through a single disciplinary lens, such as disaster response and water
insecurity challenges (Peek et al. 2020; Wutich et al. 2023).
Convergence research has become a transformative framework in scientific
research, advocating for deep collaboration and integrated solutions to
sustain scientific innovation (Baaden et al. 2024). It also serves as a
fundamental mechanism for bridging the gap between theoretical research
and practical applications (Bukvic et al. 2022; Lobo et al. 2021).
Additionally, convergence research fosters new knowledge and innovation,
driving scientific and technological advancements.

To support convergence research and scientific innovation, government
agencies such as the US National Science Foundation (NSF) and European
Union (EU) have taken initiatives on new programs. The EU's Horizon
Europe programme has a convergence strategy, where the research focuses
on tackling climate change; NSF's Growing Convergence Research and
Convergence Accelerators aim to incubate convergence across scientific
disciplines, translating basic research into solutions with real-world
applications, thus opening new frontiers in science and engineering. The
convergence process is also about the researchers themselves, as the
process of working closely with scholars from other disciplines broadens
their thinking and should provide benefits to future research.
Therefore, besides seeking advances in science and societal impacts,
these efforts also call for innovative approaches for measuring
interdisciplinary team's research convergence.

Despite the acknowledged benefits of convergence research, there are
significant challenges in measuring the effectiveness and depth of
convergence within research teams (Alwin and Hofer 2008; Feller 2006;
Petersen et al. 2021). Most related studies have typically focused on
measuring interdisciplinarity and collaboration based on scholarly
output, such as scientific publications (Rinia et al. 2002; Rinia 2007;
Raan 2005). However, due to the significant time lag in publications,
this approach has limitations in capturing the micro-dynamics of the
convergence process, hindering the ability to conduct systematic, near
real-time reflection on the team\textquotesingle s progress as the
collaboration continues. Moreover, several empirical analyses have shown
that other forms of data, such as questionnaires assessing team
collaboration, can provide insights into the convergence process (Hubbs
et al. 2020). This suggests that convergence detection can be more
timely and should incorporate more diverse data sources rather than
solely relying on publications, an area that deserves further
investigation.

To address the challenges of detecting and monitoring an
interdisciplinary team's research convergence, our research addresses a
significant gap in the existing literature by pioneering an Artificial
Intelligence (AI)-driven methodological framework that captures the
dynamic and real-time interactions among convergence participants from
various domains. We adopted and extended a value creation framework,
termed NABC (Needs, Approaches, Benefits, and Competition; Carlson and
Wilmot 2006) to practice and record the team's convergent thinking over
time. Originating in the research and development sector, this framework
has been effectively adapted for problem-solving in other areas, such as
in addressing complex environmental challenges (Westerhoff et al. 2021).
By structuring presentations and discussions around the NABC components,
research teams are compelled to articulate and align on the specific
societal needs they aim to address, the innovative approaches they will
employ, the benefits relative to existing solutions, and the competitive
landscape. NABC also provides a common structure, without any particular
disciplinary bias, to aid the convergence process.

Specifically, we developed a multi-layer, AI-driven analytical framework
to examine research convergence in interdisciplinary teams. This study
analyzes team convergence and collaboration from complementary
perspectives and addresses three research questions: (1) What viewpoints
are widely shared across the interdisciplinary team, and which remain
unique or domain-specific? (2) How do research viewpoints from different
domains influence one another within the interdisciplinary team? (3) How
do team knowledge and viewpoints progressively integrate and converge
over time?

The proposed qualitative-quantitative framework integrates large
language models (LLMs) with graph-based analytics and visualization.
LLMs are used to extract structured viewpoints and infer potential
opinion flows through carefully designed and constrained prompts, while
graph-based methods support similarity analysis, cross-domain influence
assessment, and temporal convergence analysis. To ensure reliability,
expert-level evaluation is incorporated through human-in-the-loop
validation, and analytical results are further evaluated through
cross-layer consistency among multiple analytical perspectives. To the
best of our knowledge, this work represents one of the first efforts to
provide an automated qualitative-quantitative framework for analyzing
interdisciplinary research convergence that captures both structural
relationships and longitudinal dynamics.

The remainder of this paper is organized as follows. Section 2 reviews
related literature on research convergence analysis. Section 3
introduces the interdisciplinary team and the convergence research
problem of water insecurity in underserved communities. Section 4
presents the methodological framework integrating large language models,
graph-based analytics, and human evaluation. Section 5 reports and
analyzes the results. Section 6 discusses results evaluation and the
mitigation of LLM-related uncertainty. Finally, Section 7 concludes the
paper and outlines directions for future research.

\section{\texorpdfstring{\textbf{2. Literature review}
}{2. Literature review }}\label{literature-review}

\subsection{\texorpdfstring{\textbf{2.1 Measuring research convergence
through bibliometric
analysis}}{2.1 Measuring research convergence through bibliometric analysis}}\label{measuring-research-convergence-through-bibliometric-analysis}

Many studies have attempted to measure interdisciplinarity and
collaboration, and the common approach is bibliometric analysis
(Leydesdorff and Ivanova 2021; Shi and Wang 2022). There are two
bibliometric approaches from cognitive and organizational perspectives
to analyzing publications in bibliographic databases (Abramo et al.
2018; Glänzel and Debackere 2022). They can also be combined as a hybrid
solution that covers both scholarly outputs and human factors.

\textbf{Cognitive approaches} in bibliometric analysis focus on
information flows derived from citation patterns, essential for mapping
interdisciplinary research dynamics. This method involves analyzing
linking patterns in publication data to reveal trends in collaborative
research. Notably, Rinia (2007) and Rinia et al. (2002) demonstrate the
utility of bibliometric indicators to assess the interdisciplinary
character of research and validate evaluation processes. Similarly,
Porter and Rafols (2009) highlight how citation analysis can chronicle
the degree of interdisciplinarity over time.

Expanding beyond citations, lexical or content analysis offers deeper
insights into thematic alignments and research convergence among
disciplines. Porter et al. (2006) emphasize how content analysis can
assess thematic convergence across disciplines by examining the
integration of concepts, techniques, and data. Tobi and Kampen (2018)
advocate for a combined methodology that merges content analysis with
other methods, providing a holistic view of interdisciplinary processes.
Furthermore, Huutoniemi et al. (2010) and Brink et al. (2020) suggest
integrating qualitative and quantitative assessments to evaluate
interdisciplinary research comprehensively, particularly for complex
concepts such as sustainability. Furthermore, Natural Language
Processing (NLP) is an advanced technique that has been adopted to
extract the semantics of publications. For instance, the BERTopic model
can be used to automatically find associations of interdisciplinary
classification without supervision which largely reduces manual
processing (Kim et al. 2024).

People participating in the convergence research are also important as
they are the unit practicing collaboration, and analyzing information of
people can help with decision-making and finding potential
collaborations (Boyack 2009; Goring et al. 2014; Lakhani et al. 2012;
Leahey 2018; Urbanska et al. 2019). From an organizational perspective,
the analysis of collaboration and authorship relies heavily on
researcher metadata, including experts' educational backgrounds,
institutional affiliations, and disciplinary classifications, which are
documented drivers of interdisciplinary integration and research
convergence (Aboelela et al. 2007, Carr et al. 2018, Mansilla 2006, Van
Rijnsoever and Hessels 2011). Network analysis among collaborators can
help reveal the structural aspects of convergence, for example, by
identifying central nodes (e.g., key researchers) and linkages (e.g.,
research similarity) within research communities (Li and Yu 2024).
Wagner et al. (2011) and Leydesdorff and Rafols (2011) explore various
network dynamics that are instrumental in understanding the social
dynamics that facilitate knowledge integration in interdisciplinary
research. Lungeanu et al. (2014) delve into how previous collaboration
patterns impact the formation and success of interdisciplinary
scientific teams. Additionally, Bellanca (2009) provides case studies on
the network analyses of researchers in biology and chemistry,
identifying patterns of interdisciplinary collaboration within academic
institutions.

\subsection{\texorpdfstring{\textbf{2.2. Temporal analysis of research
convergence}}{2.2. Temporal analysis of research convergence}}\label{temporal-analysis-of-research-convergence}

Monitoring the convergence of disciplines from a temporal perspective is
crucial in understanding the evolution and dynamics of interdisciplinary
research (Porter and Rafols 2019). Miyashita and Sengoku (2021)
highlight the value of temporal monitoring in their examination of
collaboration and knowledge structures within interdisciplinary
projects, demonstrating how scientometric analysis over time can reveal
the dynamic changes and complex interactions that influence the course
of interdisciplinary research. Furthermore, temporal analysis offers
insights into how the convergence of disciplines influences scientific
collaboration over extended periods. By measuring the variety and
disparity of disciplines involved, researchers can visualize the
interdisciplinary nature of research activities and assess their
evolution at both institutional and individual levels (Glänzel and
Debackere 2022). Such analysis not only highlights the dynamic nature of
interdisciplinary research but also underscores the importance of
developing methodologies that can adapt and respond to the changing
landscapes of science and technology. By integrating these temporal
dimensions, researchers can provide a more comprehensive understanding
of the mechanisms that drive successful interdisciplinary collaborations
and the long-term effects of such integrations on scientific innovation
(Zhang et al. 2021). Existing approaches analyze the evolutionary
aspects of interdisciplinary research primarily through the analysis of
bibliometric data. Although these methods offer an important perspective
that goes beyond traditional cross-sectional studies by focusing on
interdisciplinary interactions (Miyashita and Sengoku 2021), the primary
data used remain publications. As a result, they are unable to provide a
near-real-time understanding of whether a team is becoming more
convergent as the collaboration progresses.

In our research, we extend beyond the current bibliographic data based
analysis by introducing a novel methodological framework for convergence
analysis. Our proposed method integrates LLMs with network analysis and
knowledge graphs to study interdisciplinary interactions within
real-time collaborative environments. Unlike traditional methods that
rely on static, retrospective data, our approach utilizes near real-time
analysis to capture the fluid nature of interdisciplinary integration.
This methodological innovation not only enhances the accuracy and depth
of interdisciplinary research analysis but also provides a more
comprehensive toolset for tracking and assessing the evolution of these
interactions. By employing viewpoint-mining techniques and analyzing
presentations across multiple domains, our approach offers new insights
into the structural dynamics of interdisciplinary research teams. The
next section describes the convergence research problem and methodology
in detail.

\section{\texorpdfstring{\textbf{3. Convergence Research Problem for
Clean Water Access and Water
Insecurity}}{3. Convergence Research Problem for Clean Water Access and Water Insecurity}}\label{convergence-research-problem-for-clean-water-access-and-water-insecurity}

Ensuring clean water access for underserved communities is a complex
challenge that requires an integrated, cross-disciplinary approach. A
team of interdisciplinary researchers (we call it the ``Water'' team for
short), funded by the NSF Growing Convergence Research program, is
working to develop innovative solutions that combine social, physical,
and cyber-infrastructure to improve water security for underserved
\emph{colonia} communities along the U.S.-Mexico border (Wutich et al.
2022a).

This effort involves multiple disciplines working together to address
different facets of the problem. Engineers are designing customized
water purification solutions tailored to local conditions, while social
scientists and community researchers conduct interviews to understand
community needs and deploy socially suitable techniques for improving
water access (Castro-Diaz et al. 2025, Roque et al., 2024). Data
scientists leverage existing spatial data on infrastructure, water
quality, and demographics to identify which communities face the
greatest challenges. Using multi-criteria analyses, incorporating
factors such as water quality, public water service availability, and
population size, they prioritize communities most in need of
intervention. Additionally, water policy and law researchers use these
insights to engage with state and local decision-makers, advocating for
policies that address water access disparities. Addressing such a
multifaceted, real-world problem requires a convergent research paradigm
(Sundstrom et al., 2023), where disciplinary knowledge flows across
boundaries, fostering deeper cross-disciplinary understanding and
driving innovative, practical solutions.

\section{\texorpdfstring{\textbf{4.
Methodology}}{4. Methodology}}\label{methodology}

To analyze the Water team's growing collaborations, we first initiated
and documented a series of NABC presentations among team members over a
12-month period. Each presentation was required to address four aspects:
end-user and societal Needs (N), compelling Approaches (A), potential
Benefits (B) of the proposed approaches, and the superiority of these
benefits compared to the Competition (C). The transcripts of these
presentations and the associated discussions were collected and used as
the primary data source for our research convergence analysis.

We then developed a qualitative-quantitative evaluation framework
empowered by large language models, graph-based analytics, and
visualization approaches. Specifically, our research aims to understand
research convergence within our interdisciplinary team from three
complementary perspectives, framed as research questions. Figure 1
illustrates the methodological framework, in which each block represents
the workflow corresponding to one of the three interconnected research
questions:

(1) What viewpoints are widely shared across the interdisciplinary team,
and which remain unique or domain-specific?

(2) How do research viewpoints from different domains influence one
another within the interdisciplinary team?

(3) How do team knowledge and viewpoints progressively integrate and
converge over time?

To address these research questions, a core task is the systematic
extraction of viewpoints from each presentation (preprocessing module in
Figure 1). We developed a workflow that leverages large language models
(OpenAI GPT) to extract viewpoints related to the NABC aspects from
presentation transcripts. First, the LLM is provided with definitions of
the NABC framework to establish analytical context. Each transcript is
then input with instructions to identify key viewpoints (up to 10 per
presentation), summarize each viewpoint within 10 words, and label it
according to the corresponding NABC aspect. For example, from the Water
Technology Expert 1 presentation, the LLM identified two viewpoints
related to \emph{Needs}: (1) an essential need for clean and safe water
in communities, and (2) water safety concerns highlighting the
importance of quality access. Through this process, a total of 89
viewpoints were extracted from 11 presentations for subsequent analysis.
In addition, each presenter is assigned a research domain based on their
expertise, including Participatory Social Science (PSS), Water
(Treatment) Technology (WT), Water Law (WL), Community Research (CR),
Data Science (DS), and Social Science \& Hydrology (SSH).

\includegraphics[width=\textwidth]{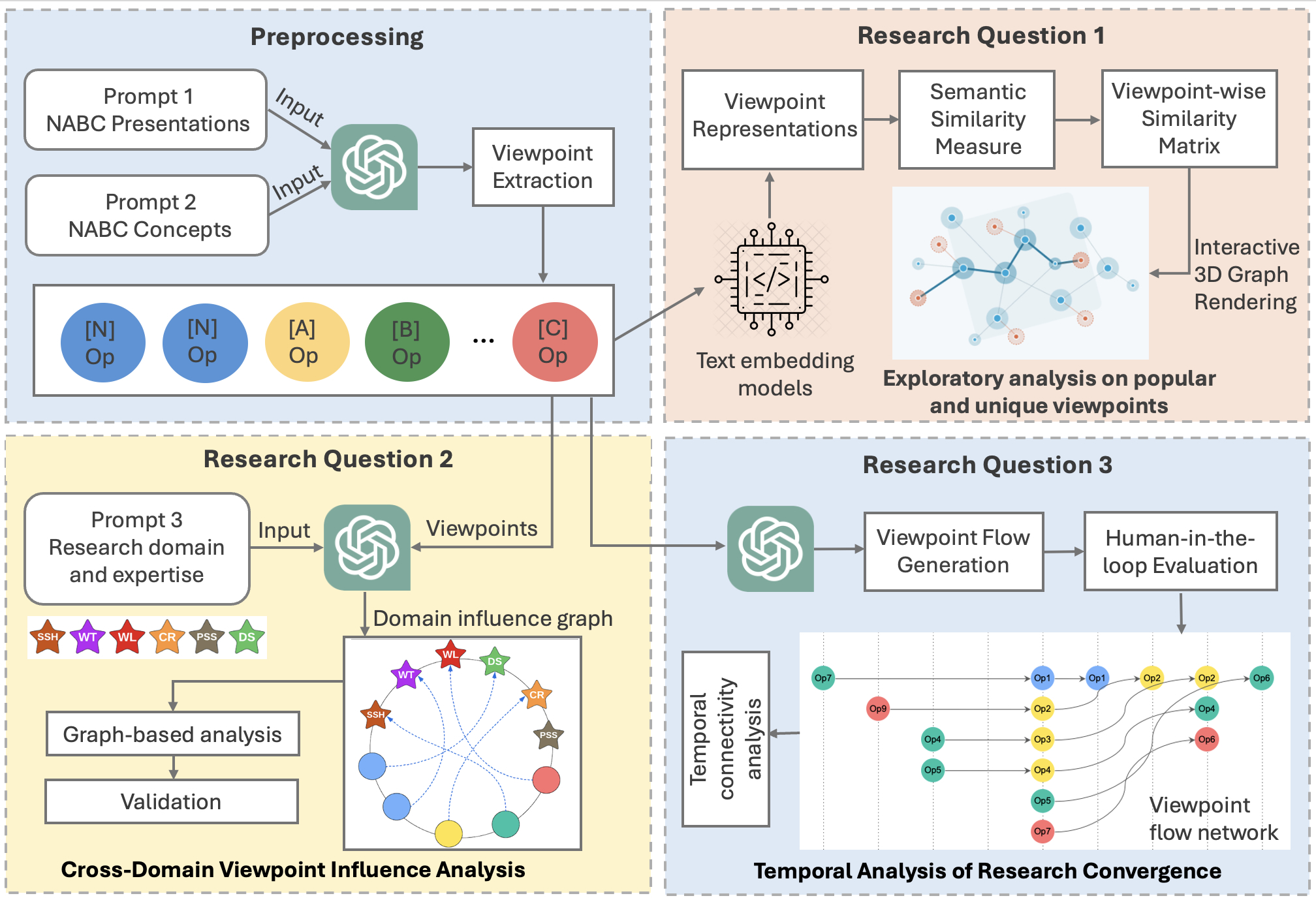}

\textbf{Figure 1}. Methodological framework for research convergence
analysis

\subsection{\texorpdfstring{\textbf{4.1 Exploratory graph visualization
to identify popular and unique
viewpoints}}{4.1 Exploratory graph visualization to identify popular and unique viewpoints}}\label{exploratory-graph-visualization-to-identify-popular-and-unique-viewpoints}

Our first research question aims to identify popular and unique
viewpoints/ideas among the team members. This is because both popular
and unique viewpoints play an important role in fostering convergent
research and collaboration as well as driving impactful publications.
Popular viewpoints create common ground, making it easier for
researchers from different domains to collaborate, build upon existing
knowledge, and co-author interdisciplinary studies. On the other hand,
unique ideas introduce innovation, challenge conventional thinking, and
open new research directions that can lead to groundbreaking
discoveries. While popular ideas facilitate immediate collaboration,
unique perspectives drive long-term advancements, making both essential
for a dynamic and influential research ecosystem.

Answering these questions advances the cross-domain analysis (described
in Section 3.2) and refines the granularity analysis at both the
individual viewpoint and domain levels. To achieve this, we first
generate a quantitative measure of similarity between viewpoints by
applying a text embedding model (i.e., text-embedding-3-small). This
model converts all textual descriptions of viewpoints into
high-dimensional vector representations. We then apply cosine similarity
to measure the semantic similarity between each viewpoint pair (Reimers
et, al. 2019). This data is subsequently loaded into our graph
visualization platform, based on the GeoGraphViz algorithm (Wang et al.
2023), to help explore the distribution patterns of viewpoints among
team members from different research domains.Traditional graph
visualization algorithms are mostly 2D, and nodes are positioned
primarily based on their connections rather than the strength of those
connections (e.g., the degree of similarity). This limits the amount of
information that can be effectively conveyed and restricts the ability
to analyze graph structures from different perspectives.

In comparison, the GeoGraphViz algorithm builds upon the concept of
force-directed graph visualization, where the placement of nodes is
influenced not only by their attractiveness but also by their
repulsiveness. In the context of our viewpoint analysis, the
attractiveness between nodes is represented by their similarity, while
their repulsiveness is indicated by their dissimilarity. Through the
combined effect of attractive and repulsive forces, the positions of
graph nodes in the visual space are dynamically adjusted until the graph
reaches an equilibrium state, where the forces among all nodes are
balanced. The resulting graph layout then reveals node distribution
patterns based on the strength of their interconnectivity (e.g.,
similarity or dissimilarity between viewpoints), which can further
support the identification of popular and unique viewpoints or potential
research ideas.

Figure 2 demonstrates the graph visualization based on viewpoint
similarity analysis. Each node represents a viewpoint extracted from the
NABC presentation, and an edge exists between two nodes when the
semantic similarity between them is above or below a threshold. Using a
selective subset of viewpoints in this analysis helps highlight the most
prominent patterns while excluding noise and less significant
connections. In addition, two nodes placed close to each other indicate
a high degree of similarity between them. Different node colors indicate
the domains of expertise from which these viewpoints originate. The
larger the node, the more connections it has to other nodes. The
analysis of this graph, in terms of viewpoint distribution patterns and
unique and shared viewpoints, can be found in Section 4.1. The graph
also allows for interactive exploration to view the 3D graph from
different angles (Figure 2b) and subgraphs (Figure 2c).

{\def\LTcaptype{none} % do not increment counter
\begin{longtable}[]{@{}
  >{\centering\arraybackslash}p{(\linewidth - 2\tabcolsep) * \real{0.4}}
  >{\centering\arraybackslash}p{(\linewidth - 2\tabcolsep) * \real{0.6}}@{}}
\endhead
\endlastfoot
\multicolumn{2}{@{}>{\centering\arraybackslash}p{(\linewidth - 2\tabcolsep) * \real{1.0000} + 2\tabcolsep}@{}}{%
\includegraphics[width=\textwidth]{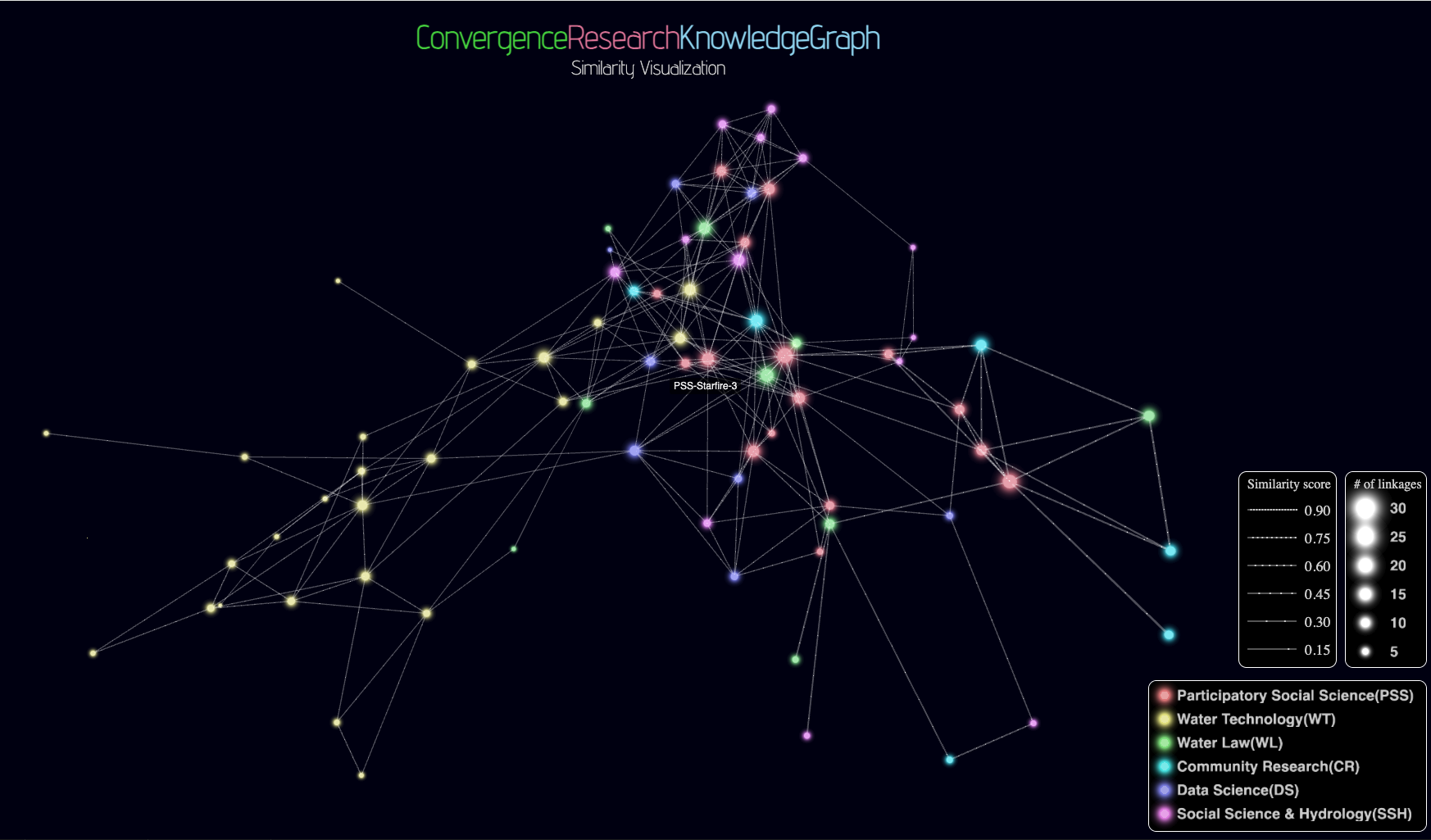}

(a)} \\
\includegraphics[width=\linewidth]{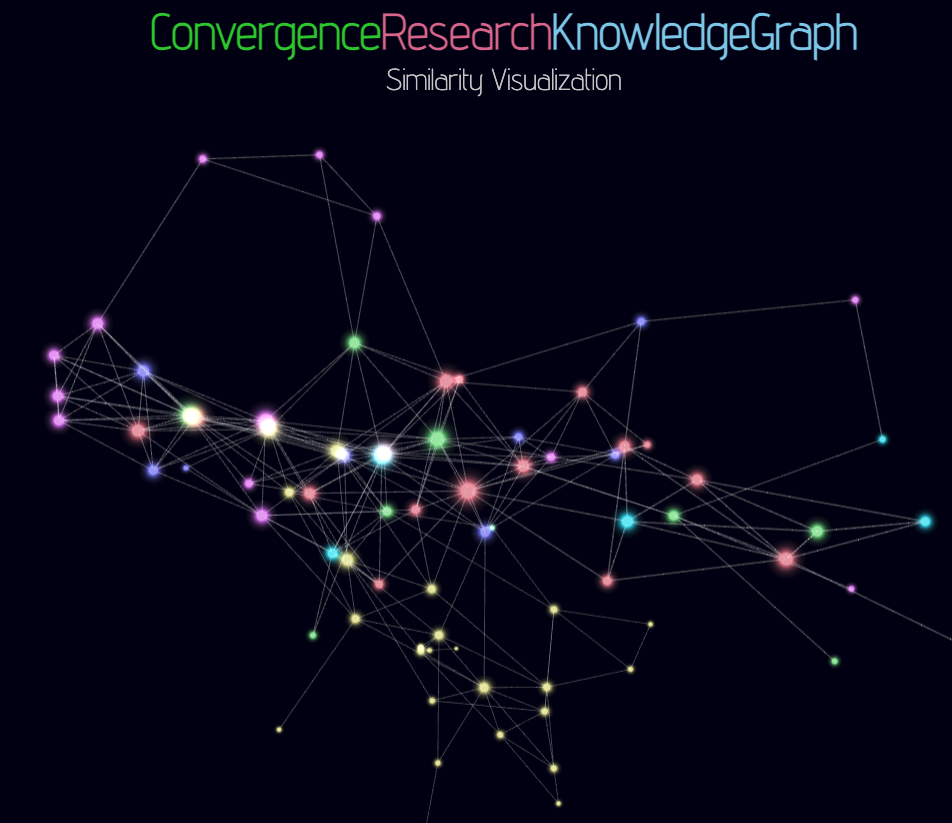}

(b) &
\includegraphics[width=\linewidth]{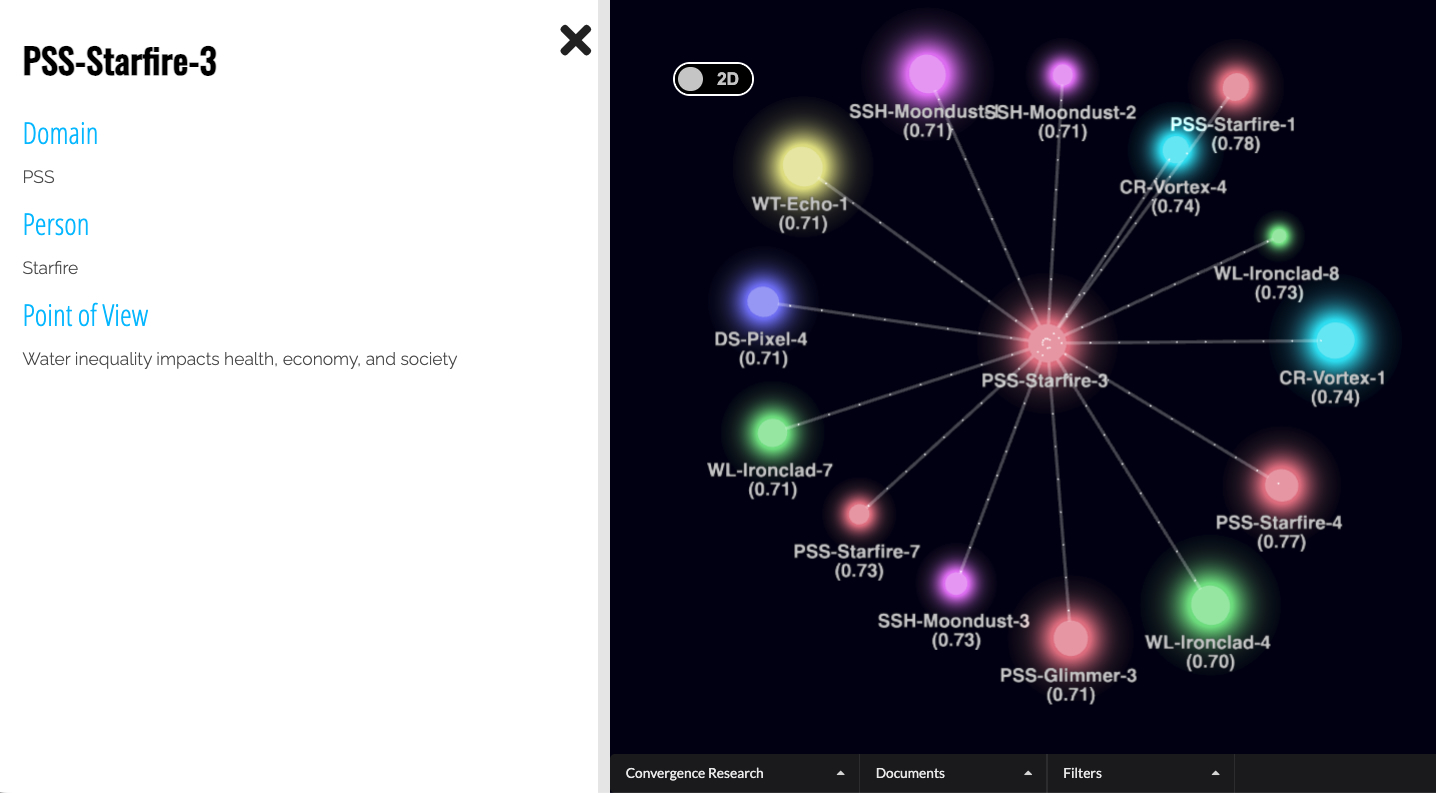}

(c) \\
\end{longtable}
}

\textbf{Figure 2}. Convergence research graph visualization based on
viewpoint similarity. Each node represents a viewpoint extracted from
the NABC presentations. Node colors indicate the domains of expertise
from which these viewpoints originate. (a) A 3D view of the similarity
graph with a similarity threshold of 0.68 (90th percentile of the
similarity value distribution); links exist between nodes whose
similarity exceeds this threshold. (b) The same graph viewed from a
different angle through interactive exploration. (c) A subgraph
connecting the node labeled ``PSS-Starfire-3,'' representing Starfire's
third viewpoint from the participatory social science (PSS) domain. The
team member's name has been anonymized to protect privacy.

It is important to note that the web-based graph visualization tool is
designed to support dynamic configurations for interactive exploration
of the 3D graph and its subgraphs. By allowing different threshold and
parameter combinations, as well as filtering techniques, the tool
enables visual examination of viewpoints and their interconnections to
help uncover new and interesting patterns.

\subsection{\texorpdfstring{\textbf{4.2 LLM and graph-based analytics to
identify cross-domain
influence}}{4.2 LLM and graph-based analytics to identify cross-domain influence}}\label{llm-and-graph-based-analytics-to-identify-cross-domain-influence}

The convergence graph visualization in Section 4.1 provides a
viewpoint-based analysis at the individual level. To further examine
cross-domain similarity and influence, we conducted a domain-level
analysis by constructing viewpoint connectivity graphs across research
domains. Specifically, we measure the similarity between a presenter's
viewpoints and each research domain, where each domain is represented by
a bag of keywords provided by its team members. An LLM is then used to
compute similarities and generate a graph linking the presenter's domain
to other research domains. Figure 3 illustrates the resulting
domain-level similarity graph derived from PSS Expert 1's presentation.
Edge labels indicate the number of viewpoints that are similar across
domains, with edge widths scaled accordingly. The goal is to identify
related or overlapping perspectives across domains and to anticipate
potential collaborations.

\includegraphics[width=\textwidth]{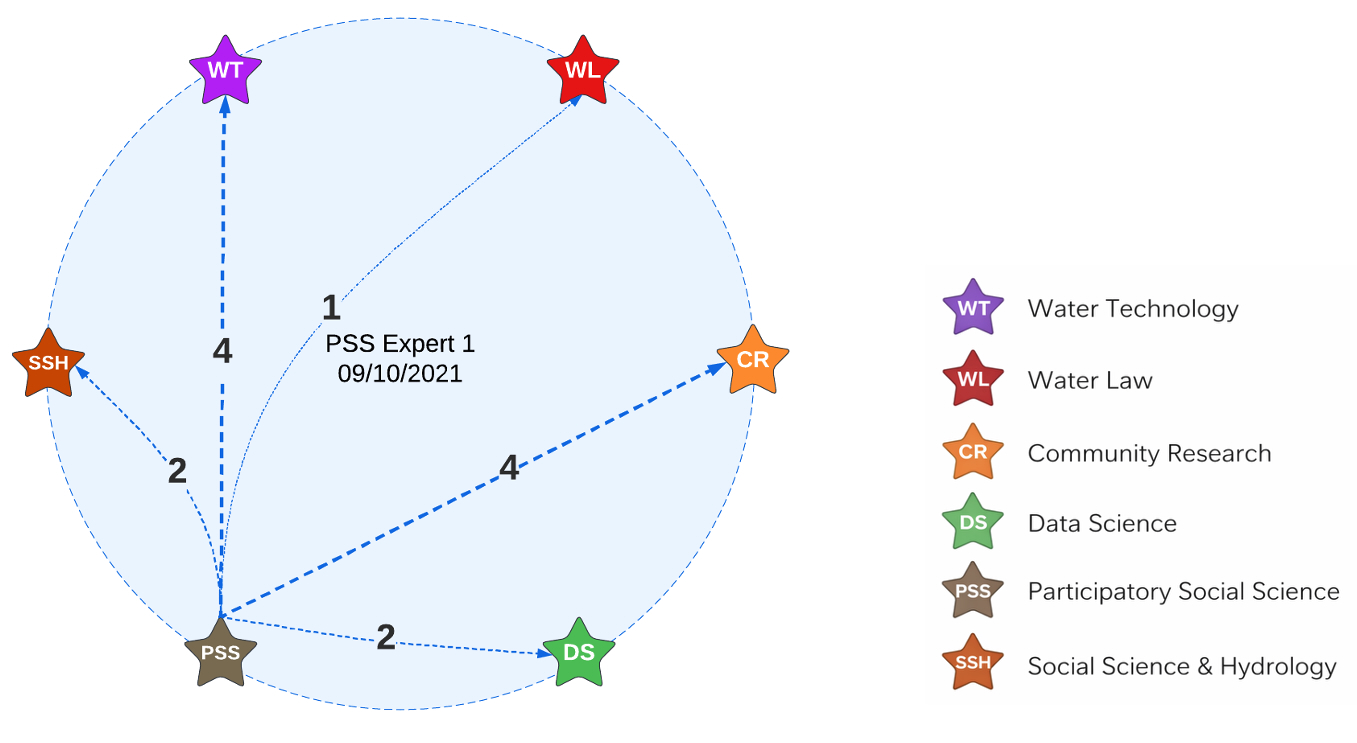}

\textbf{Figure 3}. Viewpoint connectivity graph across domains based on
the analysis of PSS Expert 1's presentation. The number on the edge
represents the number of similar viewpoints between two domains, with
the edge width adjusted accordingly.

From Figure 3, one can see that the research viewpoints of a team member
with expertise in Participatory Social Science (PSS) are highly relevant
to collaborators from Water (Treatment) Technology (WT) and Community
Research (CR), indicating a strong connection and potential
collaborations among these areas. Meanwhile, viewpoints in PSS are also
shared by a degree of 2 with team members from Social Science \&
Hydrology (SSH) and Data Science (DS), suggesting potential
collaborations among subteams in these areas as well. The shared
viewpoints between PSS and Water Law (WL) are fewer compared to the
connections with other domains for SSH.

To measure the importance of a connected domain to a presenter's domain,
Eigenvector Centrality (EC) based graph metric is applied. Eigenvector
Centrality (EC) is a measure of a node\textquotesingle s influence
within a network based not only on its direct connections but also on
the importance of the nodes it is connected to. Unlike simple degree
centrality, which counts direct links, EC assigns higher scores to nodes
that are linked to other high-scoring nodes, making it useful for
identifying key influencers in research collaborations. This method is
commonly used in ranking algorithms, such as Google\textquotesingle s
PageRank, to assess the relative importance of entities within
interconnected systems (Chandrashekhar et al. 2022). Mathematically, the
EC score can be calculated as follows.

Given a graph G with an adjacency weight matrix
\(A = \lbrack A_{i,\ j}\rbrack,\ i,j \in \{ 1,2,...,n\}\), where \(i\)
represents the presenter's domain, and \(j\) is the index of all
research domains, and \(n\) is the total number of nodes in the graph.
In the context of a viewpoint-domain connectivity graph, where the
number of nodes equals the number of domains, and two nodes have an edge
when one or more similar viewpoints are identified from a presenter's
domain \(i\) to another domain \(j\). The EC score vector
\(X = {\lbrack x_{1},\ x_{2\ }...\ x_{n}\rbrack}^{T}\) satisfies:

\begin{equation}
AX = \delta X
\end{equation}

Where \(\delta\) is an eigenvalue and \(X\) is the corresponding eigen
vector. To derive \(X\) as the EC vector, we take the value with the
largest eigenvalue \(\delta_{\max}\). Element-wise,

\begin{equation}
x_{i} = \frac{1}{\delta_{\max}} \sum_{j} A_{i,j} x_{j},
\quad \text{for } i = 1,2,\ldots,n-1
\end{equation}

The resulting EC vector \(X^{({k)}_{}}\)provides a measure of the
importance or potential influence of other domains on a speaker's
domain, denoted as \(k\). To generate an EC matrix, the EC values for
domains connected to speakers' domains are first calculated following
Equations (1)-(2). Then each pair is then formatted as \{(presenter's
domain \(k\), connected domain \(i\)): EC value \({x^{(k)}}_{i}\) and
grouped by the unique (presenter's domain, connected domain) pair. When
multiple values exist for a given domain pair, the mean value is
recorded to represent the aggregated influence and collaboration
strength or potential across domains. After iterating the EC calculation
on graphs generated from presentations in each research domain, as
illustrated in the Figure 3 example, the EC matrix connecting all
domains can be computed.

\subsection{\texorpdfstring{\textbf{4.3 Temporal analysis of research
convergence}}{4.3 Temporal analysis of research convergence}}\label{temporal-analysis-of-research-convergence-1}

To enable longitudinal analysis and to measure whether the Water team's
research perspectives and viewpoints converge over time, we have
introduced an LLM-based opinion flow generation method. The opinion flow
captures whether some team member captures another one's viewpoints and
adapts them to become a part of their viewpoints during the NABC
practices. The previous analysis described in Sections 4.2 does not
consider the time sequence in the presentations and the viewpoints
raised during the presentation, but this analysis will specifically
consider the temporal aspects of the analysis to help us understand team
convergence (or possible divergence) of research viewpoints over time.
Our method for generating opinion flows began with prompting an LLM to
analyze viewpoints extracted from a series of presentations in a given
order. The LLM was tasked with examining each presenter's viewpoints for
potential connections to those in subsequent presentations, focusing on
two types of flows: within-category flows, where viewpoints from the
same NABC category influence others, and cross-category flows, where
viewpoints across categories interact. The prompt required the LLM to
identify up to 20 flows per presenter, equally divided between
within-category and cross-category connections. For each flow, the LLM
provided reasoning that demonstrated the relevance or logical connection
between the opinions, ensuring that both thematic alignment and
progression were captured in the output. All the nodes and identified
flows form a viewpoint flow network.

To further measure the Water team's convergence over time, we analyzed
the temporal evolution of viewpoint convergence using the edge-to-node
ratio, which quantifies network connectivity per node as the viewpoint
graph grows over time. At each time step \(t\), we construct a directed
graph \(G(t) = (V(t),\ E(t))\), where nodes represent extracted
viewpoints and directed edges represent opinion flows between
presentations. The edge-to-node ratio is defined as:

\begin{equation}
r(t) = \frac{|E(t)|}{|V(t)|}
\end{equation}

, where \(|V(t)|\) is the number of nodes and \(|E(t)|\) is the number
of total edges at time \(t\ (t = 0,1,...n - 1)\), where \(n\) is the
total number of presentations.

To capture temporal dynamics, we construct a sequence of cumulative
graphs. At Time \(t = 1\), the graph includes only the first two
presentations, and for Time \(t = n - 1\), the graph incorporates all
presentations available up to that point, along with their associated
nodes and opinion flows. When a new presentation is added at time \(t\),
the graph updates as:

\begin{equation}
|V(t)| = |V(t - 1)| + n_{\text{new}}
\end{equation}

\begin{equation}
|E(t)| = |E(t - 1)| + e_{\text{new}}
\end{equation}

, resulting in an updated ratio:

\begin{equation}
r(t) = \frac{|E(t - 1)| + e_{\text{new}}}
{|V(t - 1)| + n_{\text{new}}}
\end{equation}

Beyond serving as a connectivity metric, the edge-to-node ratio provides
an indicator of viewpoint convergence. A higher ratio implies that, on
average, each opinion node participates in more opinion flows,
increasing the likelihood that opinions are repeatedly referenced,
reinforced, or adopted across presentations. This higher level of
connectivity reduces fragmentation in the opinion network and promotes
the formation of shared opinion clusters. Conversely, when new
presentations introduce nodes without corresponding edges (i.e.,
\(e_{new} = 0\)), the ratio decreases, indicating weaker integration of
new opinions and temporarily reduced convergence.

\section{\texorpdfstring{\textbf{5. Results, Analysis, and
Validation}}{5. Results, Analysis, and Validation}}\label{results-analysis-and-validation}

\subsection{\texorpdfstring{\textbf{5.1 Shared and unique viewpoints
within the interdisciplinary
team}}{5.1 Shared and unique viewpoints within the interdisciplinary team}}\label{shared-and-unique-viewpoints-within-the-interdisciplinary-team}

To identify shared and unique viewpoints across the interdisciplinary
team, we apply a 3D graph visualization. Figure 3 shows a similarity
graph in which each node represents a viewpoint, and an edge connects
two nodes when their semantic similarity exceeds a given similarity
threshold, which is determined based on visual effect and also the
similarity value distribution. As a result, the graph highlights only
closely related viewpoints rather than all extracted viewpoints, making
it easier to identify commonly shared ideas among team members.

In this graph, node size reflects the number of similar viewpoints it
connects to. Larger nodes therefore represent more popular viewpoints
shared across domains. For example, the pink node in Figure 3(a)
corresponds to a viewpoint from participant PSS-starfire (viewpoint
index 3). As shown in Figure 3(c), this viewpoint, ``Water inequality
impacts health, economy, and society,'' is similar to 14 other
viewpoints spanning all six research domains. The interactive 3D design
allows users to explore these popular viewpoints and examine their
connections within and across domains.

The overall structure of the graph reveals clear domain-level patterns.
Some domains, such as Water Technology, show strong within-domain
similarity and form relatively distinct clusters that reflect
domain-specific perspectives. Social Science and Hydrology exhibit
similar clustering patterns but with more connections to other domains.
In contrast, viewpoints from Participatory Social Sciences are more
centrally located and highly connected, highlighting their integrative
role in linking perspectives across the interdisciplinary team.

In addition to highlighting popular viewpoints, the graph can also be
configured to identify unique viewpoints that exhibit low similarity to
others. Figure 4 shows such a configuration, which can be viewed as a
dissimilarity graph. Rather than visualizing viewpoints with similarity
above a threshold, this graph includes only nodes and links where
similarity falls below a specified threshold, allowing viewpoints that
differ substantially from others to stand out.

\includegraphics[width=\textwidth]{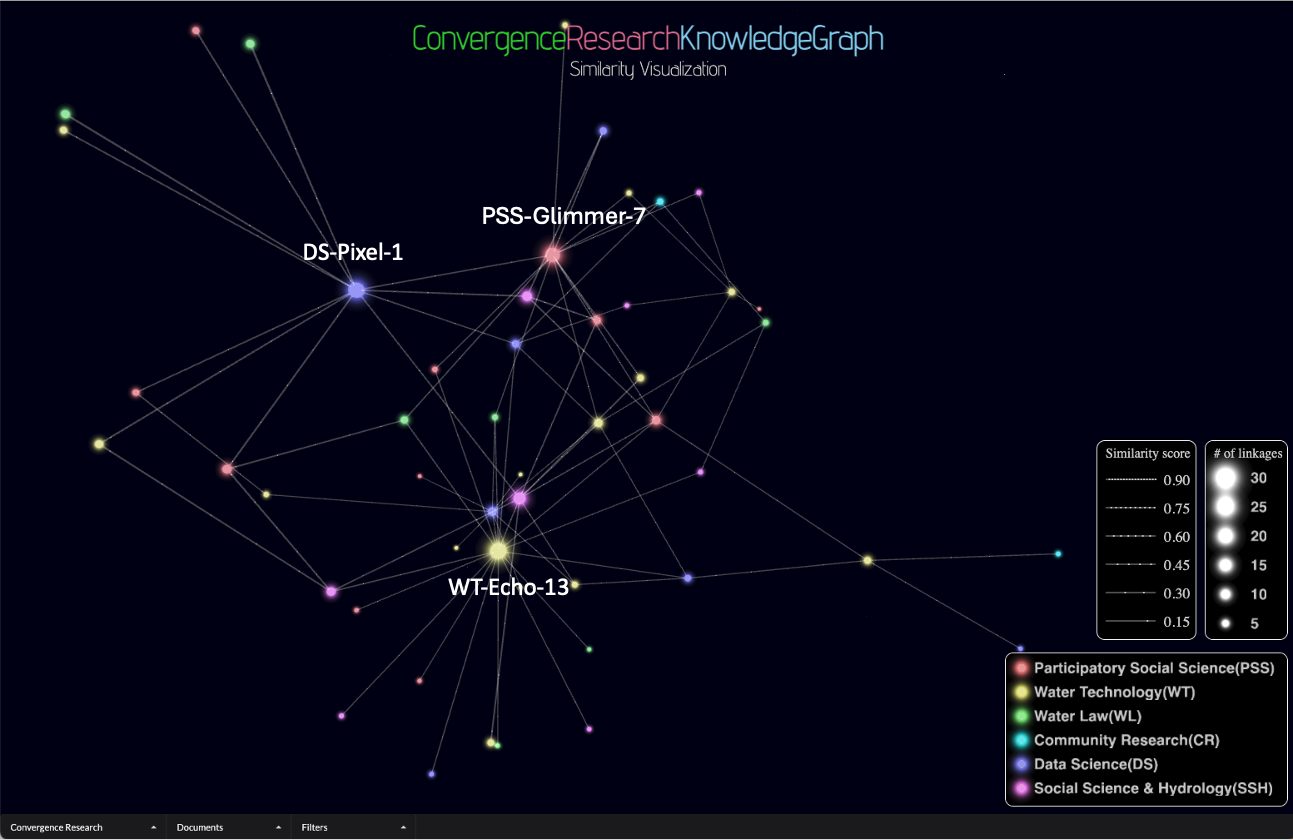}

\textbf{Figure 4}. Graph visualization for identifying unique opinions.
This graph includes only nodes (viewpoints) and links where the
similarity is below a specified semantic similarity threshold. Larger
nodes indicate viewpoints that link to more nodes under this criterion
and are therefore potentially unique viewpoints.

In Figure 4, node size indicates the number of low-similarity
connections a viewpoint has. Larger nodes therefore represent viewpoints
that are dissimilar from many others and are potentially unique. For
example, WT-Echo-13 represents a viewpoint from a Water Technology
expert stating, ``Employs sous-vide technique for efficient titanium
impregnation into carbon blocks'' (Farsad et al. 2023). This viewpoint
has 16 low-similarity connections, reflecting its highly specialized and
technical nature. Its focus on a specific material processing technique
is narrowly scoped and distinct from the broader, systems-level
discussions common across other domains, which explains its limited
overlap with other viewpoints.

Another example is PSS-Gimmer-7, a viewpoint from a Participatory Social
Science expert stating, ``Hargrove Heyman group as main competition with
a stakeholder process approach'' (Hargrove and Heyman 2021). This
viewpoint emphasizes competitive landscape analysis and stakeholder
engagement strategies rather than scientific or technical aspects of
water research. As such, it reflects domain-specific strategic knowledge
that is not widely shared across the team, making it stand out as a
unique perspective. A third example, DS-Pixel-7, presents an interesting
case. This viewpoint from a Data Science expert states, ``Shocking lack
of clean water at US-Mexico border.'' While the topic itself is well
known among water researchers, it appears as a unique viewpoint in the
graph. This is likely because data scientists may not be deeply immersed
in the domain-specific literature, and an issue that is familiar to
other team members can emerge as a newly recognized or surprising
insight from a data science perspective. As a result, this viewpoint
reflects a distinct framing shaped by disciplinary background rather
than novelty of the issue itself, illustrating how unique perspectives
can arise from differences in expertise and problem framing rather than
content alone.

\subsection{\texorpdfstring{\textbf{5.2 Cross-Domain Research
Influence}}{5.2 Cross-Domain Research Influence}}\label{cross-domain-research-influence}

Building on the qualitative insights gained from the graph
visualizations, we next examine how influence is distributed across
research domains in a more quantitative manner. Specifically, we assess
cross-domain influence on viewpoints using Eigenvector Centrality (EC),
computed from the viewpoints connectivity graph shown in Figure 3. The
resulting EC confusion matrix is presented in Figure 5. Values in each
cell indicate the influence of the presenter's domain (row labels) on
the other domains (column labels). A higher influence value suggests
that expertise from the presenter's domain is more closely linked to,
and potentially needed by, other domains in addressing the proposed
research problem. The diagonal cells represent within-domain
connectivity and are set to 1.

\includegraphics[width=\textwidth]{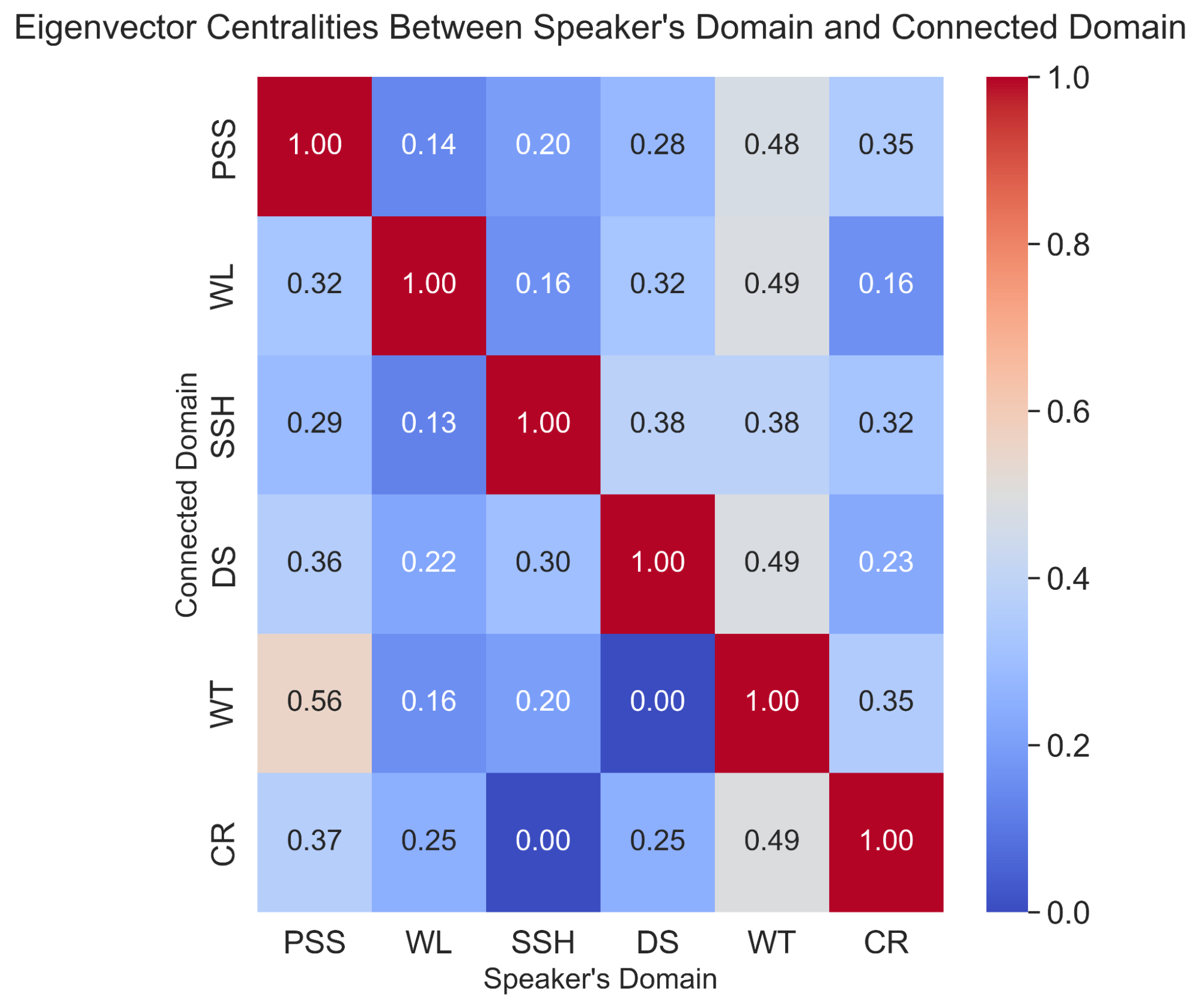}

\textbf{Figure 5}. Eigenvector centralities showing research influence
from a presenter's domain to other domains.

From Figure 5, it can be observed that Participatory Social Science
(PSS) and Water Technology (WT) are two strongly connected domains.
Regardless of whether the speaker's domain is PSS or WT, their
viewpoints are closely related, as indicated by high EC values.
Specifically, the relative influence from Water Technology to
Participatory Social Science (0.56) is slightly higher than the
influence from PSS to WT (0.48). This indicates strong familiarity and
potential for close connections between team members' research in these
two domains, suggesting strong collaboration potential. This inference
is further validated by joint publications, including Stoler et al.
(2022), Westerhoff et al. (2021), Thomson et al. (2024), and Wutich et
al. (2023), in which leads in PSS and WT and their teams collaborated
during the course of the project.

In addition, the Water Technology (WT) domain plays an important role
for all other domains within the Water team. As shown in Figure 5, when
the speaker's domain is WT (the fifth column), the corresponding EC
values are the highest compared to speakers from other domains (as shown
across columns). Specifically, the influence or potential connectivity
from Water Technology to Water Law, Data Science, and Community Research
is relatively high, with EC values of 0.49. The EC values from Water
Technology to Participatory Social Science and Community Research
follow, with values of 0.48 and 0.38, respectively. This pattern can be
explained by two main reasons. First, water quality is a central issue
in this project, and knowledge from Water (Treatment) Technology is
essential for analyzing chemical substances in drinking water. This
facilitates a more comprehensive understanding of how water quality
affects water accessibility and residents' health across other domains.
For example, Water Law experts can help protect residents' rights to
safe drinking water and health coverage. Data Science experts can
analyze the geospatial distribution of safe drinking water and propose
feasible data-driven solutions to improve access. Community Research
experts can identify wastewater challenges faced by rural communities
and investigate regionalization of water and wastewater systems.

Second, the deployment of water (treatment) technologies is
indispensable for fundamentally improving water quality and providing
colonias residents with safe drinking water. This enables Participatory
Social Science experts to conduct community-based fieldwork to
understand how individuals and communities respond to water conditions,
as well as how local environments, economies, and cultures change in
response. Social Science Hydrology experts can develop models to
understand the dynamics among public policies, water resources, and
land-use practices, which can further inform decision-making and policy
development in both the public and private sectors. These findings are
further supported by the team's joint publications, including Wutich et
al. (2022a, 2022b), Wutich et al. (2023), and Gu et al. (2023), which
are led by PSS or DS teams and involve close collaboration with the WT
team.

Finally, domain knowledge from Social Science and Hydrology (SSH) is
closely intertwined with data-driven analysis and therefore with Data
Science (DS). As shown in Figure 5, the cross-domain influence between
SSH and DS is relatively strong, with EC values of 0.38 indicating
influence from DS to SSH and 0.30 from SSH to DS. This suggests that SSH
expertise is critical for conducting accurate data-driven analyses.
Indeed, in a data-driven study led by a data scientist (Gu et al.,
2023), SSH experts were closely involved throughout the research
process. During the analysis, SSH experts communicated frequently with
data scientists, providing constructive feedback on drinking water
facilities in colonias, identifying key factors contributing to water
insecurity, and helping classify and prioritize water needs across
different underserved \emph{colonia} communities.

\subsection{\texorpdfstring{\textbf{5.3 Analysis of Water team's
research convergence over
time}}{5.3 Analysis of Water team's research convergence over time}}\label{analysis-of-water-teams-research-convergence-over-time}

This section provides further analysis of how the project's convergence
research evolved over time using temporal viewpoint flow, as shown in
Figure 6. It captures the evolution of ideas and methodologies in this
study, which focuses on converging around solutions for water security
in U.S. colonias. We map opinions extracted from several presentations
(Presentations 1--11) to show how ideas within the research project
progressed over time. Symbols along the top indicate the disciplines
represented in each presentation. Because the presentations follow the
NABC framework, each opinion (node) is color-coded to represent end-user
and societal needs (blue), approaches (yellow), benefits (green), and
benefits to the competition (red). Arrows depict the flow and influence
of these opinions across the sequence of presentations and illustrate
the dynamic and interconnected nature of this interdisciplinary research
program.

\includegraphics[width=\textwidth]{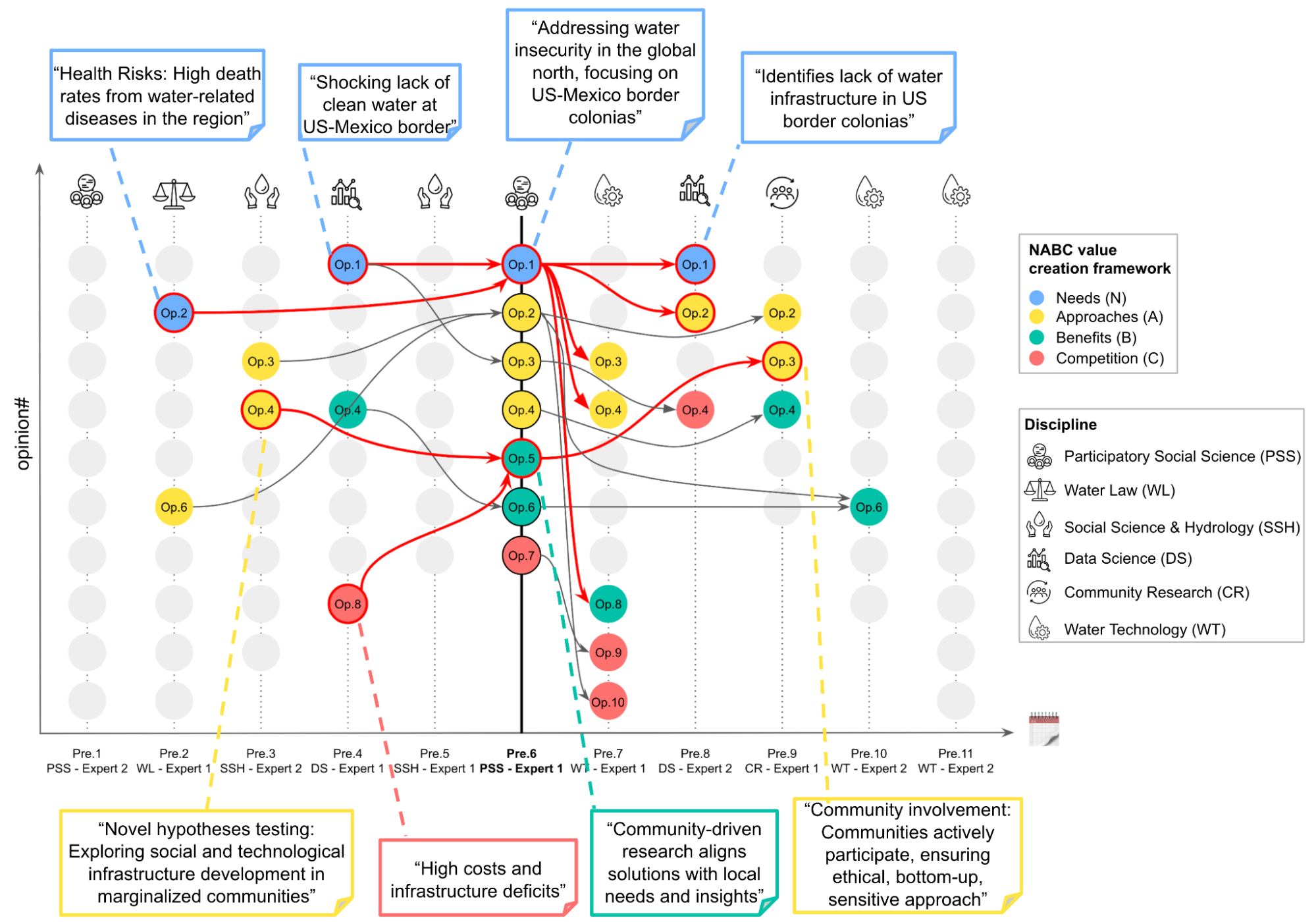}

\textbf{Figure 6}. Opinion flows across different team members and
collaborative domains. Pre: Presentation.

To examine how ideas evolve and propagate across team members over time,
we analyze opinion flow graphs, which capture how viewpoints from
earlier presentations influence those articulated in later ones. These
flow graphs make it possible to trace both within-domain and
cross-domain movements of ideas, revealing how interdisciplinary
knowledge is collectively constructed rather than developed in
isolation.

We use Presentation 6 as an illustrative example. The speaker is an
expert in Participatory Social Science (PSS-Expert 1), and Figure 6
visualizes all opinion flows associated with this presentation. From a
holistic perspective, the flow patterns show both within-category and
cross-category movements of ideas, indicating that viewpoints are shaped
by contributions from multiple domains and NABC dimensions. Two
prominent clusters of flows emerge: one related to the Need aspect,
represented by the linked blue nodes, and the other showing a mix of
NABC dimensions and research domains.

For example, two earlier opinions, Op.2 from a Water Law expert
(WL-Expert 1), stating ``Health risks: High death rates from
water-related diseases in the region,'' and Op.1 from a Data Science
expert (DS-Expert 1), stating ``Shocking lack of clean water at
US-Mexico border,'' both flow into PSS-Expert 1's Op.1, which frames the
problem as ``Addressing water insecurity in the global north, focusing
on US-Mexico border colonias.'' The inferred reasoning, generated by
LLM, suggests that these upstream opinions emphasize severe health
impacts and regional urgency, which together inform a broader and more
integrated framing of water insecurity in the same geographic context.

This reframed viewpoint then flows forward to Op.1 from another Data
Science expert (DS-Expert 2), stating ``Identifies lack of water
infrastructure in US border colonias.'' This sequence illustrates how
domain-specific observations related to health risks and infrastructure
deficits converge into a shared articulation of needs. As highlighted in
Figure 6, these four opinions, marked by blue nodes and thick red
connecting lines, all fall within the ``Needs'' category, demonstrating
how core problem definitions are reinforced and refined through
cross-domain interaction.

A second example illustrates how ideas move across NABC categories. Op.5
from PSS-Expert 1, categorized under Benefits, states ``Community-driven
research aligns solutions with local needs and insights.'' This opinion
is influenced by earlier viewpoints, including an Approach opinion
(Op.4) from a Social Science and Hydrology expert (SSH-Expert 2),
``Novel hypothesis testing: Exploring social and technological
infrastructure development in marginalized communities,'' and a
Competition opinion from a Data Science expert (DS-Expert 1), ``High
costs and infrastructure deficits.'' These upstream opinions
collectively shape a benefit-oriented framing that emphasizes community
engagement as a response to structural and economic challenges. This
benefit-oriented viewpoint then flows to Op.3 from a Community Research
expert (CR-Expert 1), which proposes involving communities directly in
decision-making processes. This progression highlights how abstract
challenges and methodological considerations are translated into
actionable approaches through interdisciplinary dialogue.

Based on the evolving viewpoint flow maps, Figure 7 further analyzed and
visualized the temporal evolution of the edge-to-node ratio in the
viewpoint network as presentations are added over time. By tracking how
this ratio changes with each new presentation, the figure provides a
quantitative view of how connectivity among viewpoints develops as the
team's research progresses. The method describing how it is calculated
can be found in Section 4.3. From Figure 7, we can observe a gradual
increase in the edge-to-node ratio over time, indicating that the
opinion network becomes increasingly connected as more presentations are
incorporated. This upward trend suggests that speakers progressively
draw on a broader and more shared set of viewpoints across domains,
rather than introducing isolated or independent opinions. As the network
grows, new viewpoints are more likely to connect to existing ones,
reflecting an increasing level of alignment and reuse of prior ideas
across presentations.

A notable exception occurs at Time 6, where the ratio decreases
following the addition of the WT-Expert 1 presentation. This
presentation introduces a new node without citing opinions from earlier
presentations, increasing the total number of nodes while leaving the
number of edges unchanged. A similar, though smaller, decrease is
observed at Time 10, corresponding to the final presentation, which also
comes from the Water Technology domain (WT-Expert 2). As identified in
earlier analyses, while viewpoints from the water technology subteam
exert strong influence on other research domains, they also exhibit
higher within-team similarity. Consequently, these presentations
generate fewer cross-presentation opinion flows, resulting in a
temporary reduction in the edge-to-node ratio. Despite these local
fluctuations, the overall increase in the edge-to-node ratio reveals a
clear trend toward a more densely connected opinion network. This
pattern indicates that the team becomes increasingly convergent over
time, with viewpoints progressively integrated, reinforced, and shared
across presentations. Rather than remaining fragmented, opinions
coalesce into interconnected clusters, suggesting a transition from
exploratory knowledge sharing toward collective sense-making and a more
coherent research direction.

\includegraphics[width=\textwidth]{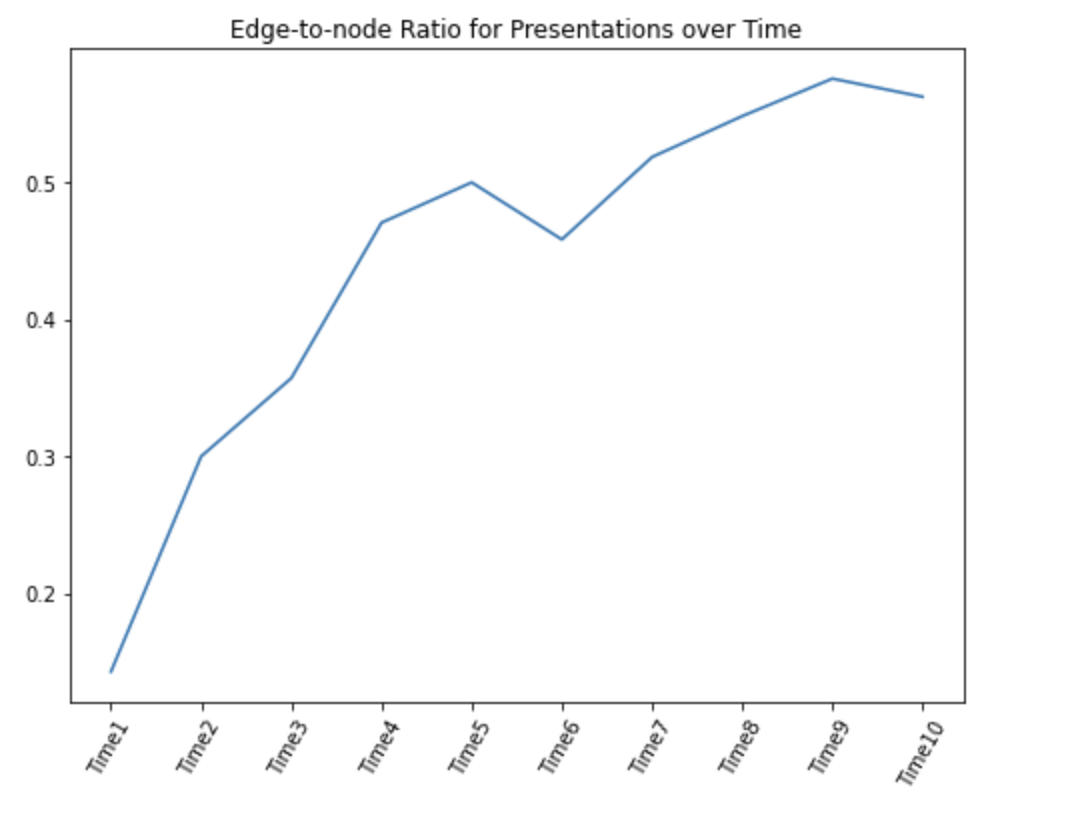}

\textbf{Figure 7}. Changes in the edge-to-node ratio of viewpoint flows
over time. The time index on the x-axis represents viewpoint flow maps
constructed using t + 1 presentations. For example, Time 1 corresponds
to a graph built from the first two presentations.

\section{\texorpdfstring{\textbf{6. Results Evaluation and Mitigation of
LLM-Related
Uncertainty}}{6. Results Evaluation and Mitigation of LLM-Related Uncertainty}}\label{results-evaluation-and-mitigation-of-llm-related-uncertainty}

This study presents a multi-layer analytical framework for examining how
interdisciplinary research teams share, integrate, and converge on ideas
over time. By combining LLMs, graph-based analysis, and visualization
with expert evaluation, the framework enables a structured and
interpretable investigation of research convergence from multiple
complementary perspectives.

While this study demonstrates the value of LLM in supporting research
convergence analysis, we explicitly recognize and address the
uncertainties introduced by LLM-based inference, including
hallucination, over-interpretation, and ambiguity in inferred
relationships. To ensure the reliability of the results, we adopt a set
of complementary evaluation and validation strategies that operate at
different stages of the analytical workflow.

First, at the viewpoint extraction stage, we mitigate LLM hallucination
by employing a grounded-inference protocol. The LLM is provided with a
rigorous definition of the NABC value creation framework to serve as a
semantic guardrail for all inferences. To ensure the reliability of the
extracted data, the prompt instructions mandate two primary constraints.
One constraint is direct textual anchoring. For every identified
viewpoint, the LLM is required to retrieve and cite supporting quotes
from the original presentation transcripts. This provenance action
ensures that viewpoints remain grounded in the experts' own words rather
than abstract summarization or free-form over-interpretation. Another
constraint is explicit mapping. Each flow is required to be a structured
triplet, i.e., \emph{(opinion1, presentor1, {[}NABC type{]})} →
\emph{(opinion2, presenter2, {[}NABC type{]})}. This prevents the model
from hallucinating connections by forcing it to map the evolution of an
idea back to a specific domain expert and a specific functional category
(e.g., mapping how a ``Need'' expressed by an environmental scientist
influences an ``Approach'' proposed by an engineer). In this way, the
risk of hallucinated opinions is mitigated by making every data point
auditable. Downstream analyses are built upon traceable viewpoints that
are explicitly linked to the source material and the specific expertise
of the research team members.

Second, for opinion flow generation, we adopt a human-in-the-loop
evaluation strategy (as shown in Figure 8) to assess the validity of
inferred viewpoint flows. Rather than treating LLM-generated flows as
ground truth, we interpret them as hypotheses about potential pathways
of idea evolution. Domain experts are invited to review each proposed
flow through a structured survey that provides the original opinion
content, the inferred direction of influence, and the reasoning
generated by the LLM. By examining both the flow direction and the
accompanying rationale, experts are able to assess logical consistency,
relevance, and interpretive accuracy at a fine level of granularity, and
to offer qualitative feedback on how well the inferred flows reflect the
evolution of ideas across presentations, with an overall disagreement
rate of approximately 7.83\%. Through this combination of
machine-generated insights and expert evaluation, the final set of
opinion flows reflects both automated reasoning and domain-specific
expertise, thereby strengthening the overall validity of the analysis.

\includegraphics[width=\textwidth]{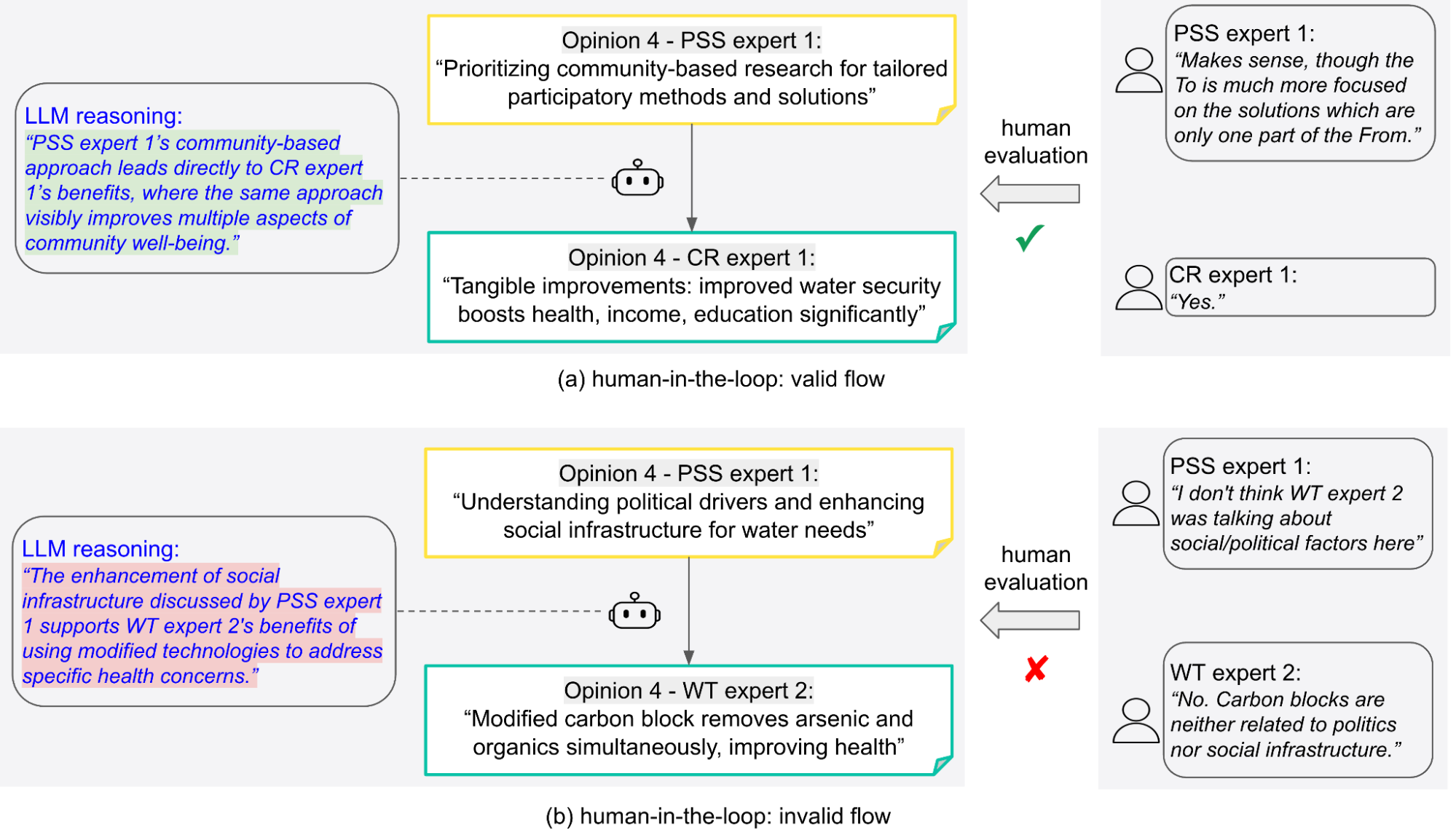}

\textbf{Figure 8}. Human-in-the-loop evaluation over viewpoint flow
generated by LLMs. CR: Community Research. PSS: Participatory Social
Science. WT: Water Technology.

Beyond explicit expert review, result evaluation is further strengthened
through cross-layer consistency among analytical outcomes. Patterns
identified through similarity-based qualitative analysis align with
quantitative measures of cross-domain influence, while both are
consistent with the temporal structures revealed by opinion flow
analysis. For example, all three analytical aspects show that Water
Technology, while having high interaction and collaboration with other
domains such as Participatory Social Science, demonstrates a higher
degree of within-domain viewpoint similarity than the other domains.
This mutual reinforcement across analytical layers provides an
additional, implicit form of validation, increasing confidence that the
observed convergence patterns are robust rather than artifacts of any
single method or model.

Taken together, these evaluation strategies ensure that LLM-enabled
analysis in this study is both transparent and reliable. By combining
prompt-level constraints, expert-driven validation, and cross-layer
consistency checks, the framework mitigates LLM-related uncertainty
while preserving the scalability and exploratory benefits of LLMs. This
approach demonstrates how LLMs can be responsibly integrated into
analytical pipelines for interdisciplinary research evaluation, with
careful attention to result credibility and methodological rigor.

In addition to addressing analytical reliability, we also place strong
emphasis on protecting the privacy and confidentiality of the
presentation scripts used in this study. All LLM-based analyses were
conducted using the OpenAI Enterprise platform, which offers
institutional data protection guarantees, including that submitted data
are not used for model training or retained beyond the scope of
authorized processing. This deployment ensures that sensitive discussion
notes and presentation content remain secure and compliant with
institutional data governance policies. By combining technical
safeguards at the platform level with constrained prompting and human
oversight at the analytical level, the framework supports responsible
use of LLMs while respecting the privacy of interdisciplinary research
discussions.

\section{\texorpdfstring{\textbf{7.
Conclusion}}{7. Conclusion}}\label{conclusion}

This study presents a new LLM-based analytical framework for examining
research convergence within interdisciplinary teams by integrating
qualitative visualization, quantitative influence analysis, and
exploratory opinion flow analysis. Together, these methods provide
complementary perspectives on how viewpoints are expressed, connected,
and evolve across domains.

Our analysis highlights the importance of both widely shared and
domain-specific viewpoints and shows that they can be systematically
identified and examined using similarity-based visualization to provide
insight into collective alignment as well as intellectual
distinctiveness within the team. These different types of perspectives
contribute in complementary ways to interdisciplinary progress. Broadly
shared viewpoints help establish cognitive alignment and coordination
across fields, while more distinctive perspectives expand the conceptual
landscape and stimulate novel directions of inquiry. The dynamic
interplay between convergence and differentiation is therefore central
to sustaining both collaborative productivity and intellectual
advancement.

The cross-domain influence analysis further reveals that domains with
stronger connectivity and shared viewpoints tend to form closer
collaborative relationships, which is reflected in the emergence of
subteams and joint research outputs. We also find that the NABC
framework provides a common structure for framing diverse perspectives
and organizing ideas, and these structured viewpoints serve as essential
input for analyzing research convergence and interdisciplinary dynamics.

Overall, this work demonstrates how combining structured collaboration
practices with multi-layer analytical methods can make interdisciplinary
research processes more visible and interpretable. The proposed
framework offers a novel AI-driven methodological approach for studying
how interdisciplinary teams converge over time and how collaborative
structures and research directions co-develop.

\subsection{\texorpdfstring{\textbf{Acknowledgements}}{Acknowledgements}}\label{acknowledgements}

The work is in part supported by the National Science Foundation under
award GCR 2021147. The authors thank Zhining Gu for support in the
analysis and visualization.

\subsection{\texorpdfstring{\textbf{References}}{References}}\label{references}

\begin{enumerate}
\def\labelenumi{\arabic{enumi}.}
\item
  Aboelela, Sally W., Jacqueline A. Merrill, Kathleen M. Carley, and
  Elaine Larson. "Social network analysis to evaluate an
  interdisciplinary research center." \emph{Journal of Research
  Administration} 38, no. 1 (2007): 61-75.
\item
  Abramo, Giovanni, Ciriaco Andrea D'Angelo, and Lin Zhang. "A
  comparison of two approaches for measuring interdisciplinary research
  output: The disciplinary diversity of authors vs the disciplinary
  diversity of the reference list." \emph{Journal of Informetrics} 12,
  no. 4 (2018): 1182-1193.
\item
  Alwin, Duane F., and Scott M. Hofer\. "Opportunities and challenges
  for interdisciplinary research." \emph{Handbook of cognitive aging:
  Interdisciplinary perspectives} (2008): 1.
\item
  Baaden, Philipp, Michael Rennings, Marcus John, and Stefanie
  Bröring. "On the emergence of interdisciplinary scientific
  fields:(how) does it relate to science convergence?." \emph{Research
  Policy} 53, no. 6 (2024): 105026.
\item
  Bellanca, Leana. "Measuring interdisciplinary research: analysis
  of co-authorship for research staff at the University of York."
  \emph{Bioscience Horizons} 2, no. 2 (2009): 99-112.
\item
  Boyack, Kevin W. "Using detailed maps of science to identify
  potential collaborations." \emph{Scientometrics} 79, no. 1 (2009):
  27-44.
\item
  Brink, Matthijs, Geerten M. Hengeveld, and Hilde Tobi.
  "Interdisciplinary measurement: A systematic review of the case of
  sustainability." \emph{Ecological Indicators} 112 (2020): 106145.
\item
  Bukvic, Anamaria, Kyle Mandli, Donovan Finn, Talea Mayo, Gabrielle
  Wong-Parodi, Alexis Merdjanoff, Joshua Alland et al. "Advancing
  interdisciplinary and convergent science for communities: lessons
  learned through the NCAR early-career faculty innovator program."
  \emph{Bulletin of the American Meteorological Society} 103, no. 11
  (2022): E2513-E2532.
\item
  Carr, Gemma, Daniel P. Loucks, and Günter Blöschl. "Gaining
  insight into interdisciplinary research and education programmes: A
  framework for evaluation." \emph{Research Policy} 47, no. 1 (2018):
  35-48.
\item
  Castro-Diaz, Laura, Anais Roque, Amber Wutich, Laura Landes,
  WenWen Li, Rhett Larson, Paul Westerhoff et al. "Participatory
  convergence: Integrating convergence and participatory action
  research." \emph{Minerva} 63, no. 3 (2025): 535-555.
\item
  Carlson, Curtis R., and William W. Wilmot. \emph{Innovation: The
  five disciplines for creating what customers want}. Currency, 2006.
\item
  Chandrashekhar, Suvarna Saumya, Mashrin Srivastava, B. Jaganathan,
  and Pankaj Shukla. "PageRank Algorithm using Eigenvector
  Centrality-\/-New Approach." \emph{arXiv preprint arXiv:2201.05469}
  (2022).
\item
  Farsad, Alireza, Mariana Marcos-Hernandez, Shahnawaz Sinha, and
  Paul Westerhoff. "Sous vide-inspired impregnation of amorphous
  titanium (hydr) oxide into carbon block point-of-use filters for
  arsenic removal from water." \emph{Environmental science \&
  technology} 57, no. 48 (2023): 20410-20420.
\item
  Feller, Irwin. "Multiple actors, multiple settings, multiple
  criteria: issues in assessing interdisciplinary research."
  \emph{Research evaluation} 15, no. 1 (2006): 5-15.
\item
  Finn, Donovan, Kyle Mandli, Anamaria Bukvic, Christopher A. Davis,
  Rebecca Haacker, Rebecca E. Morss, Cassandra R. O'Lenick et al.
  "Moving from interdisciplinary to convergent research across
  geoscience and social sciences: Challenges and strategies."
  \emph{Environmental Research Letters} 17, no. 6 (2022): 061002.
\item
  Glänzel, Wolfgang, and Koenraad Debackere. "Various aspects of
  interdisciplinarity in research and how to quantify and measure
  those." \emph{Scientometrics} 127, no. 9 (2022): 5551-5569.
\item
  Goring, Simon J., Kathleen C. Weathers, Walter K. Dodds, Patricia
  A. Soranno, Lynn C. Sweet, Kendra S. Cheruvelil, John S. Kominoski,
  Janine Rüegg, Alexandra M. Thorn, and Ryan M. Utz. "Improving the
  culture of interdisciplinary collaboration in ecology by expanding
  measures of success." \emph{Frontiers in Ecology and the Environment}
  12, no. 1 (2014): 39-47.
\item
  Gu, Zhining, Wenwen Li, Michael Hanemann, Yushiou Tsai, Amber
  Wutich, Paul Westerhoff, Laura Landes et al. "Applying machine
  learning to understand water security and water access inequality in
  underserved colonia communities." \emph{Computers, Environment and
  Urban Systems} 102 (2023): 101969.
\item
  Hargrove, William L., and Josiah M. Heyman. "A comprehensive
  process for stakeholder identification and engagement in addressing
  wicked water resources problems." \emph{Land} 9, no. 4 (2020): 119.
\item
  Hubbs, Graham, Michael O\textquotesingle Rourke, and Steven Hecht
  Orzack, eds. \emph{The toolbox dialogue initiative: the power of
  cross-disciplinary practice}. CRC Press, 2020.
\item
  Huutoniemi, Katri. "Evaluating interdisciplinary research."
  \emph{The Oxford handbook of interdisciplinarity} 10 (2010): 309-320.
\item
  Kim, Keungoui, Dieter F. Kogler, and Sira Maliphol. "Identifying
  interdisciplinary emergence in the science of science: combination of
  network analysis and BERTopic." \emph{Humanities and Social Sciences
  Communications} 11, no. 1 (2024): 1-15.
\item
  Lakhani, Jahan, Karen Benzies, and K. Alix Hayden. "Attributes of
  interdisciplinary research teams: A comprehensive review of the
  literature." \emph{Clinical and Investigative Medicine} 35, no. 5
  (2012): E260-E265.
\item
  Leahey, Erin. "The perks and perils of interdisciplinary
  research." \emph{European Review} 26, no. S2 (2018): S55-S67.
\item
  Leydesdorff, Loet, and Inga Ivanova. "The measurement of
  ``interdisciplinarity'' and ``synergy'' in scientific and
  extra‐scientific collaborations." \emph{Journal of the Association for
  Information Science and Technology} 72, no. 4 (2021): 387-402.
\item
  Leydesdorff, Loet, and Ismael Rafols. "Indicators of the
  interdisciplinarity of journals: Diversity, centrality, and
  citations." \emph{Journal of informetrics} 5, no. 1 (2011): 87-100.
\item
  Li, Jing, and Qian Yu. "Scientists' disciplinary characteristics
  and collaboration behaviour under the convergence paradigm: A
  multilevel network perspective." \emph{Journal of Informetrics} 18,
  no. 1 (2024): 101491.
\item
  Lobo, José, Marina Alberti, Melissa Allen-Dumas, Luís MA
  Bettencourt, Anni Beukes, Luis A. Bojórquez Tapia, Wei-Qiang Chen et
  al. "A convergence research perspective on graduate education for
  sustainable urban systems science." \emph{npj Urban Sustainability} 1,
  no. 1 (2021): 39.
\item
  Lungeanu, Alina, Yun Huang, and Noshir S. Contractor.
  "Understanding the assembly of interdisciplinary teams and its impact
  on performance." \emph{Journal of informetrics} 8, no. 1 (2014):
  59-70.
\item
  Mansilla, Veronica Boix. "Assessing expert interdisciplinary work
  at the frontier: an empirical exploration." \emph{Research evaluation}
  15, no. 1 (2006): 17-29.
\item
  Miyashita, Shuto, and Shintaro Sengoku. "Scientometrics for
  management of science: Collaboration and knowledge structures and
  complexities in an interdisciplinary research project."
  \emph{Scientometrics} 126, no. 9 (2021): 7419-7444.
\item
  Peek, Lori, Jennifer Tobin, Rachel M. Adams, Haorui Wu, and Mason
  Clay Mathews. "A framework for convergence research in the hazards and
  disaster field: The natural hazards engineering research
  infrastructure CONVERGE facility." \emph{Frontiers in Built
  Environment} 6 (2020): 110.
\item
  Petersen, Alexander M., Mohammed E. Ahmed, and Ioannis Pavlidis.
  "Grand challenges and emergent modes of convergence science."
  \emph{Humanities and Social Sciences Communications} 8, no. 1 (2021):
  1-15.
\item
  Petersen, Alexander Michael, Felber Arroyave, and Ioannis
  Pavlidis. "Methods for measuring social and conceptual dimensions of
  convergence science." \emph{Research Evaluation} 32, no. 2 (2023):
  256-272.
\item
  Porter, Alan, and Ismael Rafols. "Is science becoming more
  interdisciplinary? Measuring and mapping six research fields over
  time." \emph{scientometrics} 81, no. 3 (2009): 719-745.
\item
  Porter, Alan L., J. David Roessner, Alex S. Cohen, and Marty
  Perreault. "Interdisciplinary research: meaning, metrics and nurture."
  \emph{Research evaluation} 15, no. 3 (2006): 187-195.
\item
  Reimers, Nils, and Iryna Gurevych. "Sentence-BERT: Sentence
  Embeddings Using Siamese BERT-Networks." In \emph{Proceedings of the
  2019 Conference on Empirical Methods in Natural Language Processing
  and the 9th International Joint Conference on Natural Language
  Processing} (2019): 3982--92. Hong Kong, China: Association for
  Computational Linguistics.
  \\mbox{\href{https://www.google.com/search?q=https://doi.org/10.18653/v1/D19-1410&authuser=1}{\ul{https://doi.org/10.18653/v1/D19-1410}}}.
\item
  Rinia, Ed, Thed van Leeuwen, and Anthony van Raan. "Impact
  measures of interdisciplinary research in physics."
  \emph{Scientometrics} 53, no. 2 (2002): 241-248.
\item
  Rinia, Eduard Jan. "Measurement and evaluation of
  interdisciplinary research and knowledge transfer." PhD diss., Leiden
  University, 2007.
\item
  Roque, A., Wutich, A., Brewis, A., Beresford, M., Landes, L.,
  Morales-Pate, O., ... \& Water Equity Consortium, A. F. (2024).
  Community-based participant-observation (CBPO): A participatory method
  for ethnographic research. \emph{Field Methods}, \emph{36}(1), 80-90.
\item
  Shi, Yong, and Yunong Wang. "Research on the rules of scientists'
  knowledge sharing based on interdisciplinary measurement."
  \emph{Procedia Computer Science} 199 (2022): 657-664.
\item
  Stoler, Justin, Wendy Jepson, Amber Wutich, Carmen A. Velasco,
  Patrick Thomson, Chad Staddon, and Paul Westerhoff. "Modular,
  adaptive, and decentralised water infrastructure: promises and perils
  for water justice." \emph{Current Opinion in Environmental
  Sustainability} 57 (2022): 101202.
\item
  Sundstrom, Shana M., David G. Angeler, Jessica G. Ernakovich,
  Jorge H. García, Joseph A. Hamm, Orville Huntington, and Craig R.
  Allen. "The emergence of convergence." \emph{Elem Sci Anth} 11, no. 1
  (2023): 00128.
\item
  Thomson, Patrick, Justin Stoler, Amber Wutich, and Paul
  Westerhoff. "MAD water (modular, adaptive, decentralized) systems: New
  approaches for overcoming challenges to global water security."
  \emph{Water security} 21 (2024): 100166.
  \href{https://www.google.com/search?q=https://doi.org/10.1016/j.wasec.2024.100166&authuser=1}{\ul{https://doi.org/10.1016/j.wasec.2024.100166}}.
\item
  Tobi, Hilde, and Jarl K. Kampen. "Research design: the methodology
  for interdisciplinary research framework." \emph{Quality \& quantity}
  52, no. 3 (2018): 1209-1225.
\item
  Urbanska, Karolina, Sylvie Huet, and Serge Guimond. "Does
  increased interdisciplinary contact among hard and social scientists
  help or hinder interdisciplinary research?." \emph{PLoS One} 14, no. 9
  (2019): e0221907.
\item
  Van Raan, Anthony FJ. "Measurement of central aspects of
  scientific research: Performance, interdisciplinarity, structure."
  \emph{Measurement: Interdisciplinary Research and Perspectives} 3, no.
  1 (2005): 1-19.
\item
  Van Rijnsoever, Frank J., and Laurens K. Hessels. "Factors
  associated with disciplinary and interdisciplinary research
  collaboration." \emph{Research policy} 40, no. 3 (2011): 463-472.
\item
  Wagner, Caroline S., J. David Roessner, Kamau Bobb, Julie Thompson
  Klein, Kevin W. Boyack, Joann Keyton, Ismael Rafols, and Katy Börner.
  "Approaches to understanding and measuring interdisciplinary
  scientific research (IDR): A review of the literature." \emph{Journal
  of informetrics} 5, no. 1 (2011): 14-26.
\item
  Wang, Sizhe, Wenwen Li, and Zhining Gu. "GeoGraphViz:
  Geographically constrained 3D force‐directed graph for knowledge graph
  visualization." \emph{Transactions in GIS} 27, no. 4 (2023): 931-948.
\item
  Westerhoff, Paul, Amber Wutich, and Curt Carlson. "Value
  propositions provide a roadmap for convergent research on
  environmental topics." \emph{Environmental Science \& Technology} 55,
  no. 20 (2021): 13579-13582.
\item
  Wutich, Amber, Wendy Jepson, Carmen Velasco, Anais Roque, Zhining
  Gu, Michael Hanemann, Mohammed Jobayer Hossain et al. "Water
  insecurity in the Global North: A review of experiences in US colonias
  communities along the Mexico border." \emph{Wiley Interdisciplinary
  Reviews: Water} 9, no. 4 (2022a): e1595.
  \href{https://doi.org/10.1002/wat2.1595}{\ul{https://doi.org/10.1002/wat2.1595}}.
\item
  Wutich, Amber, Asher Rosinger, Alexandra Brewis, Melissa
  Beresford, Sera Young, and Household Water Insecurity Experiences
  Research Coordination Network. "Water sharing is a distressing form of
  reciprocity: Shame, upset, anger, and conflict over water in twenty
  cross‐cultural sites." \emph{American Anthropologist} 124, no. 2
  (2022b): 279-290.
  \href{https://doi.org/10.1111/aman.13682}{\ul{https://doi.org/10.1111/aman.13682}}.
\item
  Wutich, Amber, Patrick Thomson, Wendy Jepson, Justin Stoler,
  Alicia D. Cooperman, James Doss‐Gollin, Anish Jantrania et al. "MAD
  water: Integrating modular, adaptive, and decentralized approaches for
  water security in the climate change era." \emph{Wiley
  Interdisciplinary Reviews: Water} 10, no. 6 (2023): e1680.
  \href{https://doi.org/10.1002/wat2.1680}{\ul{https://doi.org/10.1002/wat2.1680}}.
\item
  Zhang, Lin, Beibei Sun, Lidan Jiang, and Ying Huang. "On the
  relationship between interdisciplinarity and impact: Distinct effects
  on academic and broader impact." \emph{Research Evaluation} 30, no. 3
  (2021): 256-268.
\end{enumerate}

\end{document}